\documentclass[%
 aip,
  pof,     
 amsmath,amssymb,
 reprint,%
 groupedaddress,
]{revtex4-1}

\usepackage{graphicx}
\usepackage{dcolumn}
\usepackage{bm}
\usepackage{verbatim}
\usepackage{multirow}
\usepackage{booktabs}
\usepackage{color}

\begin{document}

\title{Temperature Dependence of Inertial Pumping in Microchannels}

\author{Pavel E. Kornilovitch}
 \email{pavel.kornilovich@hp.com}
\affiliation{HP Inc., Corvallis, Oregon 97330 USA}  

\author{Tyler Cochell}
 \affiliation{HP Inc., Corvallis, Oregon 97330 USA} 

\author{Alexander N. Govyadinov}
 \affiliation{HP Inc., Corvallis, Oregon 97330 USA} 
  

\date{\today}  

\begin{abstract}

Inertial pumping is a promising new method of moving fluids through microchannels but many of its properties remain unexplored. In this work, inertial flow rates are investigated for different channel lengths, operating temperatures, and resistor pulse energies. The flow in closed channels is visualized by adding fluorescent tracer beads to the test fluid (pure water). A robust methodology of extracting flow rates from high-resolution video recordings is developed. Flow rates are found to scale inversely with the channel length. The observed dependence is explained based on a simple phenomenological ``kick'' model of inertial pumping. Flow rates are also fitted to the more fundamental one-dimensional model of inertial pumping from which the intrinsic drive bubble strength is extracted. The measured flow rates vary strongly with temperature. For well-developed drive bubbles, flow rates at $T = 70\,^\circ\!{\rm C}$ are about 12$\times$ higher than at $T = 30\,^\circ\!{\rm C}$. Three separate effects contribute to increasing flow rates at high temperatures: (i) lower viscosity of the test fluid, (ii) a stronger drive bubble, and (iii) increasing mechanical efficiency of the pump, i.e., better conversion of the drive bubble strength to unidirectional post-collapse kick. Relative contributions of the three effects are quantified. The energy dependence of flow rates exhibits a clear saturation behavior. The bubble strength is fitted to a phenomenological saturation model. In the end, a complete predictive length-temperature-energy model of flow rates is constructed. The observed strong temperature dependence of inertial pumping should be considered when designing microfluidic workflows. It also highlights the need for integrated flowmeters that could stabilize complex flow patterns via sensory feedback.        

\end{abstract}


\maketitle

\section{\label{tdep:sec:one}
Introduction
}

Inertial micropumps are an emerging microfluidic technology capable of effectively pumping fluids in 10-500 $\mu$m-wide channels without need for any external pressure source.~\cite{Yuan1999,Yin2005a,Yin2005b,Torniainen2012,Govyadinov2016} Based on the commercial thermal inkjet (TIJ) technology,~\cite{Stasiak2012} inertial micropumps can be fabricated in high densities up to tens of thousands per square centimeter. They can drive extensive fluidic networks performing complex workflows using only electrical power and can eventually become a key enabler of the true ``lab-on-a-chip,'' as opposed to ``chip-in-a-lab,'' devices. Physically, an inertial pump is a resistive microheater (thermal inkjet, or TIJ, resistor) located inside a closed microfluidic channel. A microsecond-long current pulse vaporizes a thin interfacial layer of fluid and creates a vapor {\em drive bubble}, which performs mechanical work through expansion. If the resistor is placed asymmetrically between two reservoirs, the expansion-collapse dynamics of the drive bubble results in a unidirectional flow from the shorter to the longer side of the channel. Physics behind pump operation is complex.~\cite{Yuan1999,Torniainen2012,Govyadinov2016,Kornilovitch2013} A key role is played by vortices formed in reservoirs,~\cite{Torniainen2012} which dissipate fluid's momentum and break the expansion-collapse symmetry. Momentum dissipation proceeds faster on the shorter side of the channel. As a result, the shorter side refills faster and arrives at collapse with a higher mechanical momentum, which translates into residual post-collapse flow. Transition from expansion to refill is delayed on the longer side due to a larger mass of fluid, hence, the term ``inertial pumping.''        

Inertial pumping was first theorized~\cite{Yuan1999} in 1999 and later demonstrated~\cite{Yin2005a,Yin2005b} by Prosperetti. In 2009, the Hewlett-Packard (HP) group observed~\cite{Torniainen2012} inertial pumping in microfluidic devices fabricated on the same manufacturing platform as HP's commercial printheads. A phenomenological reduced-parameter one-dimensional model of inertial pumping was developed in Ref.~\onlinecite{Kornilovitch2013}. It was based on the original work by Yuan and Prosperetti.~\cite{Yuan1999} The model was recently extended to two pumps acting simultaneously by Hayes et al.~\cite{Hayes2021} Careful analysis of single-pulse dynamics using a million-frame-per-second imaging technique~\cite{Govyadinov2016} confirmed basic understanding of the pumping mechanism and validated both phenomenological and first-principle CFD treatments of the effect. Multiple pumps working within a branched fluidic network were demonstrated in Ref.~\onlinecite{Hayes2018}.

\begin{figure*}[t]
\includegraphics[width=0.90\textwidth]{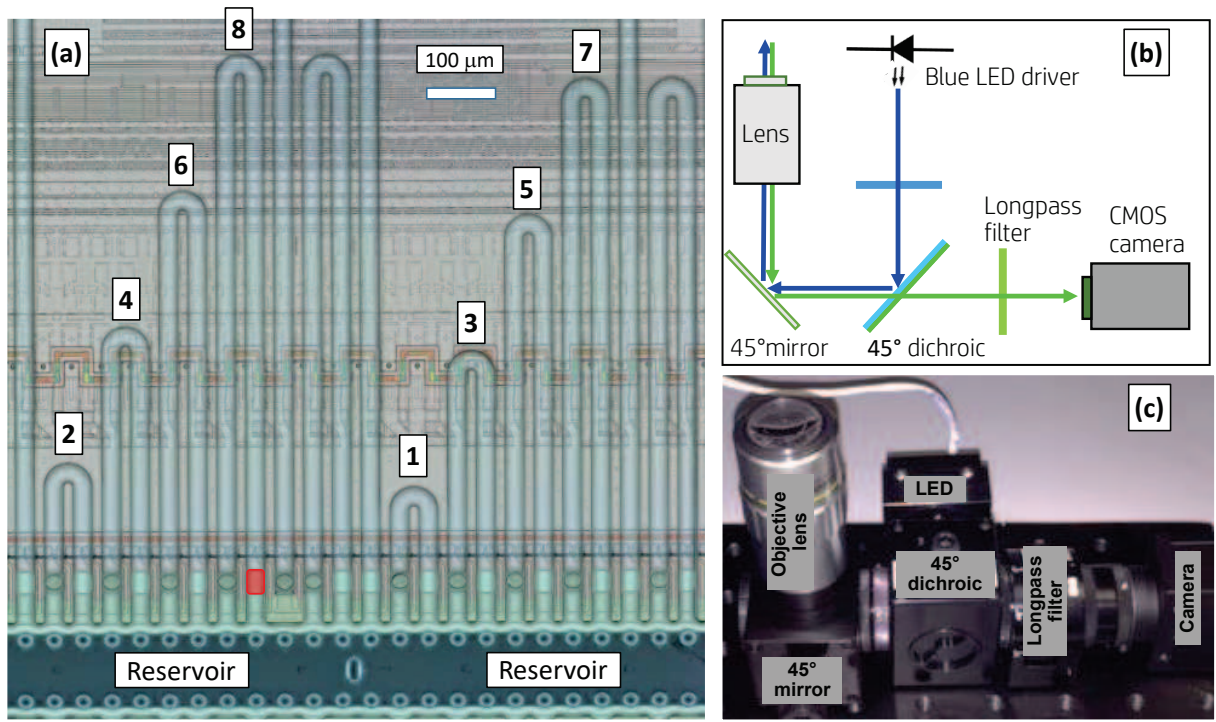}
\caption{(a) Fluidic architecture. Channel lengths are given in Table~\ref{tdep:tab:one}. The red rectangle marks the inertial pump of channel 8. The ``reservoir'' is a through-silicon slot that delivers test fluid from the back of the die. The reservoir acts simultaneously as an inlet and outlet when the fluid circulates in the anticlockwise direction. The scalebar size is 100 $\mu$m. (b) Optical diagram. (c) Photograph of the optical system. }   
\label{tdep:fig:one}
\end{figure*}

Being inertial in nature, the pumping is expected to work better at {\em high} Reynolds numbers, at least up to centimeter-wide channels. At those dimensions, instantaneous boiling is inefficient and thermal actuation needs to be replaced with mechanical impacts or other means. Inertial pumping was also demonstrated with laser~\cite{Wang2004,Dijkink2008} actuation. However, it is the low-Re limit that makes inertial pumping with thermal actuation stand out. As channels become smaller, the resistor width decreases proportionally, keeping the bubble power per die surface area constant. (Peak pressures inside drive bubbles reach tens of bars.) However, since the channel height also decreases, bubble power per {\em cross-sectional} area of the channel goes {\em up}. This positive scaling partially offsets a fourth-power increase in channel's fluidic resistance and keeps inertial pumping effective down to channel widths of 10 $\mu$m and less. Magnified by pulse frequencies of tens of kilohertz, even a sub-picoliter-per-pulse pump can deliver flow rates sufficient for many applications.       

Many physical properties of inertial pumping remain uncharacterized. In this work, we systematically investigate variation of flow rates with channel length $L$, operating temperature $T$, and resistor pulse energy $E$. We develop a robust method of extracting flow rates from raw videos based on identifying the fastest tracer particle in Sec.~\ref{tdep:sec:twothree}. To interpret results, we derive a simple phenomenological kick model of inertial pumping, which is appropriate for long channels in Sec.~\ref{tdep:sec:fourtwo}. Experimentally, we find flow rates to scale $\propto 1/L$, consistent with a $\propto L$ increase in fluidic resistance. Temperature variation of flow rates is found to be significant. Between $T = 30\,^{\circ}\!{\rm C}$ and $T = 70\,^{\circ}\!{\rm C}$ flow rates of pure water increase by as much as 12$\times$. The flow rates rise sharply because of (i) falling viscosity, (ii) stronger drive bubbles, and (iii) improving mechanical efficiency of the pump. By fitting experimental data to phenomenological models and extracting intrinsic bubble strengths, we are able to quantify the relative contributions of all three mechanisms. This is done in Sec.~\ref{tdep:sec:four}. The energy dependence of flow rates shows a clear saturation behavior indicating transition to fully developed drive bubbles. In the end, a complete predictive Length-Temperature-Energy model of inertial pumping is constructed. Our conclusions are presented in Sec.~\ref{tdep:sec:five}.

\section{\label{tdep:sec:two}
System and Methods 
}

\subsection{\label{tdep:sec:twoone}
Fluidic architecture 
}

Microfluidic devices were fabricated using the same manufacturing processes as HP's commercial inkjet printheads.~\cite{Stasiak2012} The starting substrates were 200 mm silicon wafers with CMOS driving electronics. Microheaters (TIJ resistors) and metal contacts were fabricated by thin-film deposition and patterning techniques. The fluidic layer was then formed on top of resistors using a multilayer SU8 deposition and patterning process. Finally, slots were etched through the silicon substrate to enable fluid delivery from the back of the die.   

The final architecture is shown in Fig.~\ref{tdep:fig:one}(a). It includes series of eight U-shape channels connected to one common fluid reservoir. The channel width is 30 $\mu$m and channel height is 31 $\mu$m. Channel lengths are listed in Table~\ref{tdep:tab:one}. Each channel is equipped with an inertial pump -- a $20 \times 30$ $\mu$m$^2$ resistive heater -- in the east leg of the channel. When the resistor starts firing, fluid begins to circulate through the channel in an anticlockwise direction.

\begin{table}[b]
\renewcommand{\tabcolsep}{0.2cm}
\renewcommand{\arraystretch}{1.5}
\begin{tabular}{|c|c||c|c|}
\hline\hline
    Channel     &    Length, $\mu$m      &     Channel    & Length, $\mu$m    \\ 
\hline\hline 
       1        &       396              &        5       &   1196            \\ \hline
       2        &       466              &        6       &   1266            \\ \hline
       3        &       796              &        7       &   1596            \\ \hline
       4        &       866              &        8       &   1666            \\ \hline 
\hline 
\end{tabular}
\caption{
Total lengths of the U-shape channels studied.  
} 
\label{tdep:tab:one}
\end{table}

\subsection{\label{tdep:sec:twotwo}
Experiment
}

The test fluid was pure de-ionized water. To visualize flow, 2.0 $\mu$m diameter fluorescent tracer beads (Poly\-bead\textsuperscript{\textregistered} Microspheres, Polysciences, Inc.) were added. Bead density was adjusted to have between one and ten beads in a channel at any given time, so that the tracers could be considered Lagrangian. The top SU8 layer (channel ``ceiling'') was optically transparent which allowed for easy observation of the tracers. The pumps were driven by electric firing pulses 1.0 $\mu$s in duration, with a total heat energy released in a single pulse varied between 0.896 and 1.186 $\mu$J. The firing frequency was varied between 10 and 50 Hz. Each firing created a vapor drive bubble that resulted in a unidirectional flow pulse via the inertial pumping mechanism.~\cite{Torniainen2012,Kornilovitch2013} Individual flow pulses lasted less than 400 $\mu$s for all the conditions studied in this work. Thus, short pulses were separated by long (20-100 ms in duration) waits in between. Since the pulses were independent, the overall flow rate was simply the net-flow-per-pulse times the firing frequency. Conversely, net-flow-per-pulse, which is the basic intrinsic property of the pump, could be obtained by dividing average flow rate by the firing frequency.

\begin{figure}[t]
\includegraphics[width=0.48\textwidth]{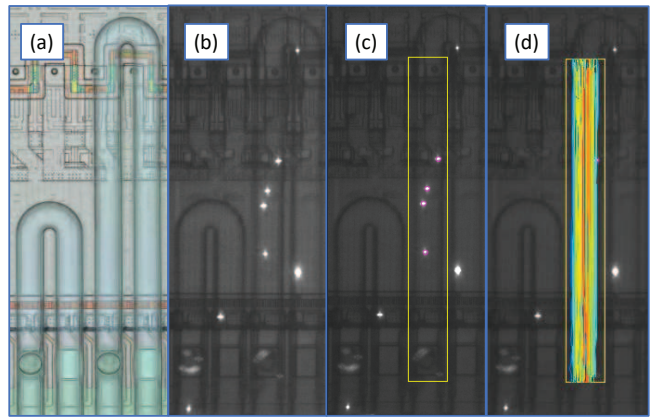}
\caption{Particle tracking sequence. (a) Dry microchannels. (b) Channels filled with water. The bright white spots are tracer particles added to water to visualize flow. (c) Particle tracking software Fiji TrackMate~\cite{Tinevez2017} identifies tracer spots within a region of interest (yellow rectangle) and marks them with magenta circles. (d) By linking spots from different video frames, the software constructs particle tracks. In this image, the tracks are colored according to their mean velocity. Faster tracers are grouped near the center of the channel, which is expected in developed laminar flow.}   
\label{tdep:fig:two}
\end{figure}

Our microfluidic chips included a built-in tempera\-tu\-re control mechanism comprising FET heaters, tem\-perature-sensitive resistors (TSRs), and CMOS circuitry that implemented a digital feedback loop. Die temperature could be specified in the control software. Although the TSRs returned temperature near the bottom of fluidic channels, computer modeling indicated that temperature variation throughout the channel thickness (31 $\mu$m) was less than $1\,^{\circ}\!{\rm C}$ and could be neglected for our purposes. Experimentally, die temperature was varied between $T = 30\,^{\circ}\!{\rm C}$ and $T = 70\,^{\circ}\!{\rm C}$, but some results were also collected at $T = 80\,^{\circ}\!{\rm C}$. Characterizing temperature dependence of inertial pumping was the main goal of this investigation.    

The optical system utilized to observe moving tracers was similar to the one described in Ref.~\onlinecite{Hayes2018}. The optical diagram is shown in Fig.~\ref{tdep:fig:one}(b) and real setup in Fig.~\ref{tdep:fig:one}(c). Observations were performed using a custom-designed inverted optical microscope with 10$\times$ magnification. The basic design described earlier~\cite {Torniainen2012} was modified to enable fluorescent imaging capability by adding a set of filters and dichroic beam splitter using illumination from a Lightspeed Technologies HPLS-36DD18B series single emitter with a built-in driver for strobing a 478 nm blue LED. The frame rate of the recording camera was synchronized with the pumps' pulse rate, typically at a 1:1 ratio. Thus, tracers' displacement observed over, say, 200 video frames was the result of 200 consecutive pump pulses. Video records were processed using Fiji software package~\cite{Schindelin2012} and Fiji TrackMate plugin.~\cite{Tinevez2017} The result was a series of tracer tracks and their respective mean velocities in pixels-per-frame as shown in Fig.~\ref{tdep:fig:two}. The region of interest was limited to straight channel segments, which ensured that tracers moved primarily along the $y$ axis and their $x$ coordinates were approximately constant. Brownian motion of 2.0 $\mu$m tracers was unobservably small in our experimental setup. The mean velocity of each track was, therefore, interpreted as an estimate for the flow field $v(x)$ at the track's $x$ position in the channel. Both $x$ and $v$ were converted from pixels to micrometers using a known scale factor. The number of tracks extracted from raw video records varied between 100 and 1000 for different experimental conditions. Such statistics enabled reconstruction of the entire flow profile and subsequent calculation of average flow rates with sufficient accuracy as explained in the next section.

\begin{figure*}[t]
\includegraphics[width=0.98\textwidth]{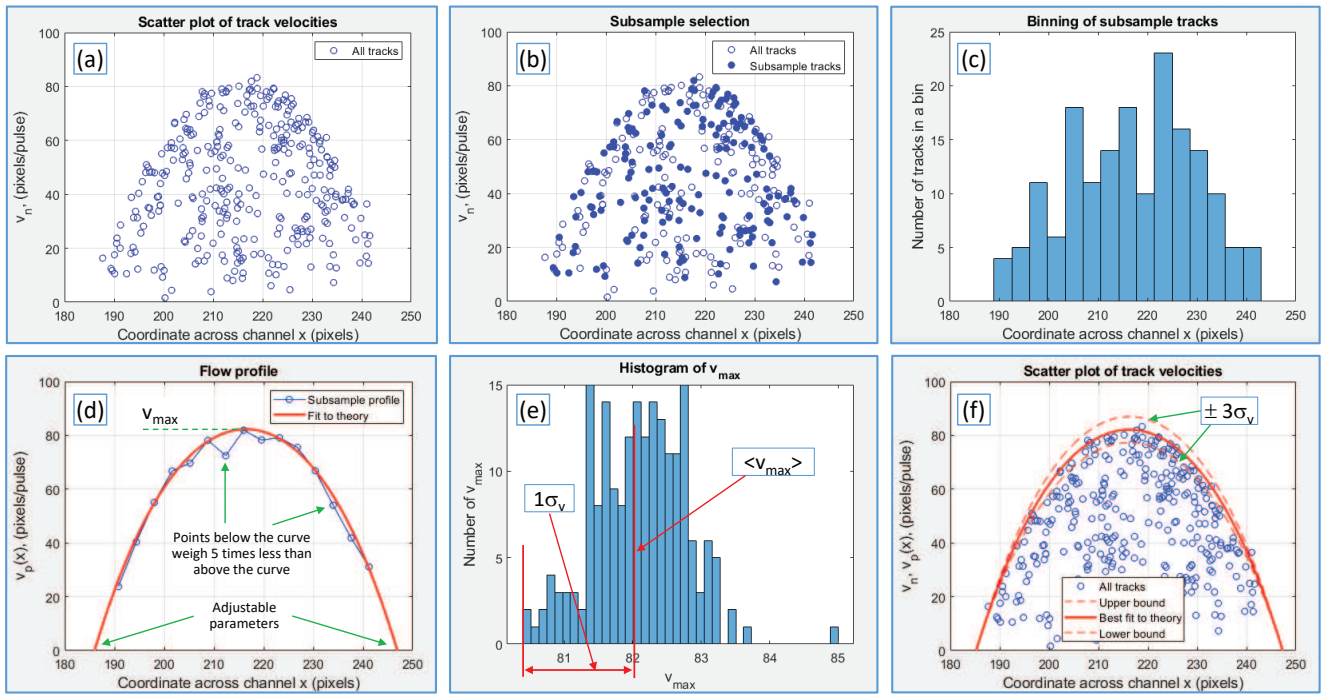}
\caption{Flow rate measurement workflow illustrated on the loop 4, $T = 70\,^{\circ}\!{\rm C}$, $E = 1.053$ $\mu$J data set. (a) The starting point: a scatterplot of tracer velocities vs position in the channel produced by Fiji TrackMate,~\cite{Tinevez2017} after removal of outliers. (b) Half of tracks are randomly selected to form a subsample. (c) The sabsample tracks are grouped into 15 bins of equal width. The number of bins $K$ should be chosen carefully. If $K$ is too large, there will be too few tracks in each bin and the subsample flow profile will fluctuate wildly. However, a small $K$ does not resolve the flow profile enough, which reduces quality of the subsequent fit. (d) Within $k$-th bin, a maximum velocity $v_p(x_k)$ is computed, which is then taken as an estimate of the flow profile at the bin's midpoint $x_{k}$. Collectively, $v_p(x_k)$ forms a subsample flow profile. Then, the set of pairs $\{ x_k ; v_p(x_k) \}$ is fitted to the theoretical profile of Eqs.~(\ref{flnet:eq:atwentyfourthree}) and (\ref{flnet:eq:atwentyfourfour}) using the $x$ coordinates of the two zeros and the overall height as three adjustable parameters. The height of the optimum pseudo-parabola is the $v_{\rm max}$ of the given subsample. Note that during optimization, we used an asymmetric price function: points below the theoretical curve were priced five times less than points above the curve. This is because the points below the curve may result from insufficient data (a bin may be missing a tracer with velocity close enough to the true profile, which artificially depressed the sample profile at the corresponding $x_k$) while the points above the curve may not. (e) Steps (b)-(d) are repeated 200 times producing a distribution of $v_{\rm max}$. The mean value of the distribution is taken as the final estimate of the true maximal velocity, and half-width as one standard deviation. (f) The original scatterplot of (a) superimposed on the best-fit theoretical profile as well as profile's upper and lower bounds. }   
\label{tdep:fig:three}
\end{figure*}

\subsection{\label{tdep:sec:twothree}
Flow rate measurements 
}

Once a set of tracks is identified, one way to determine an overall flow rate is to average tracks' velocities. This method implicitly assumes that tracers are distributed uniformly across the channel cross section. However, because of several physical effects, this assumption is never valid. First, due to their finite size, the tracers cannot get too close to the channel's walls, so the low-velocity parts of the flow field are undersampled. Second, the tracers can settle or cluster. They also interact with the walls. Finally, the tracers are disturbed by the drive bubble when flow over the pump, which moves them across the channel in an unpredictable manner. Figure~\ref{tdep:fig:three} shows that even a high-statistics set of tracks clearly lacks sufficient uniformity to be sure that simple averaging will produce an accurate result. Additionally, it is not easy to assess the cross-sectional distribution experimentally, as the tracers' $z$ coordinates remain unresolved in our setup.

We found that a more robust method of calculating the overall flow rate relies on identifying the {\em fastest tracer}. In a well-developed Poiseuille flow, there exists a one-to-one correspondence between the maximal and average velocities. For the simplest cross-sectional shapes, the ratio of the two is known theoretically. For rectangular channels, the ratio is given in Appendix, Eqs.~(\ref{flnet:eq:atwentyfourone})-(\ref{flnet:eq:atwentythreetwo}). For 30/31 channels studied in this work, $v_{\rm max}/\langle v \rangle = 2.096$, see Fig.~\ref{flnet:fig:eleven}(a).       

There is a legitimate question about applicability of fully deve\-loped flow profiles to pulsatory flows generated by inertial pumps. All our prior studies based on published~\cite{Torniainen2012,Govyadinov2016} and unpublished CFD simulations have shown that net fluid displacement after an inertial pulse matches a fully developed flow profile with deviations less than 1\% at any point of the cross-section. Although the topic deserves further investigation, the two flow types can be considered indistinguishable for practical purposes, as statistical variations result in standard deviations of order 2\%-3\% as shown below.

\begin{figure*}[t]
\includegraphics[width=0.98\textwidth]{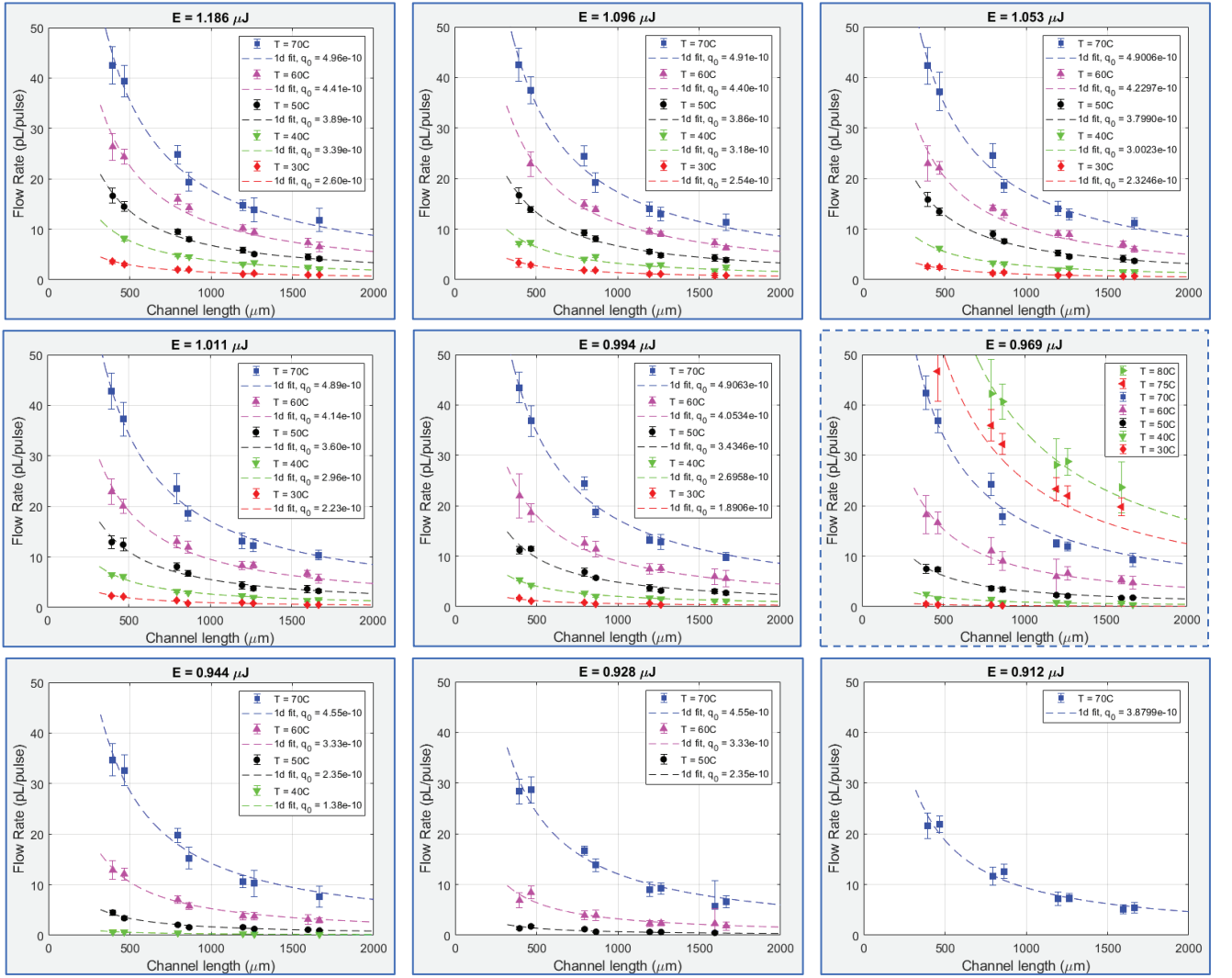}
\caption{Inertial flow rates in pL/pulse vs channel length $L$. Each panel corresponds to a particular resistor energy and mul\-tiple curves within a panel correspond to different operating temperatures. The symbols are mean experimental values, and error bars are three times the standard deviation computed by the resampling procedure. Dividing flow rates by the channel cross-sectional area of 930 $\mu$m$^2$ yields an average linear displacement per pulse. For example, a flow rate of 40 pL/pulse corresponds to an average linear displacement of 43 $\mu$m/pulse. The dashed lines are best fits to the 1d model of inertial pumping, see Sec.~\ref{tdep:sec:fourone}. Phenomenological bubble strengths extracted from the fits are given in the legends in units of kg$\,$m/s. The panel with a dashed outline ($E = 0.969$ $\mu$J) also includes data for $T = 75\,^{\circ}\!{\rm C}$ and $80\,^{\circ}\!{\rm C}$. }   
\label{tdep:fig:four}
\end{figure*}

The calculation workflow is illustrated in Fig.~\ref{tdep:fig:three}. Once a set of tracks $\{v_n\}$ is identified and cleaned from outliers, Fig.~\ref{tdep:fig:three}(a), taking the max operation is all that is needed in the case of large and dense data sets. Then, the flow rate is simply
\begin{equation}
Q {\rm [pL/pulse]} = \frac{{\rm max}\{v_n\}[\mu{\rm m/pulse]} \cdot A[\mu{\rm m}^2]}
{2.096 \cdot 1000 [\mu{\rm m}^3/{\rm pL} ] } \: ,
\label{tdep:eq:one}
\end{equation}
where $A = 30 \times 31 = 930$ $\mu$m$^2$ is the cross-sectional area and the coefficient 1000 in the denominator converts cubic micrometers to picoliters that are more natural volume units for our system. This simple procedure is justified by an expectation that within a large set of tracks, there is {\em always} one that is sufficiently close to the channel center, and the sample maximum is a good estimator of a true maximum velocity.      

The described process does not work very well at low statistics when the presence of a tracer near the channel center is not guaranteed. The max operation might also be skewed by unobvious outliers near $v_{\rm max}$. The method's robustness can be improved further by considering the entire {\em flow profile} $v_{\rm max}(x)$, that is the maximal flow velocity at a given point $x$ across the channel. The function $v_{\rm max}(x)$ is also known theoretically for any channel aspect ratio, see Appendix, Eqs.~(\ref{flnet:eq:atwentyfourthree}) and (\ref{flnet:eq:atwentyfourfour}). Anchoring the side arms of the profile enables robust determination of the profile's pinnacle even in the absence of well-defined tracks near the maximal velocity.

\begin{figure*}[t]
\includegraphics[width=0.98\textwidth]{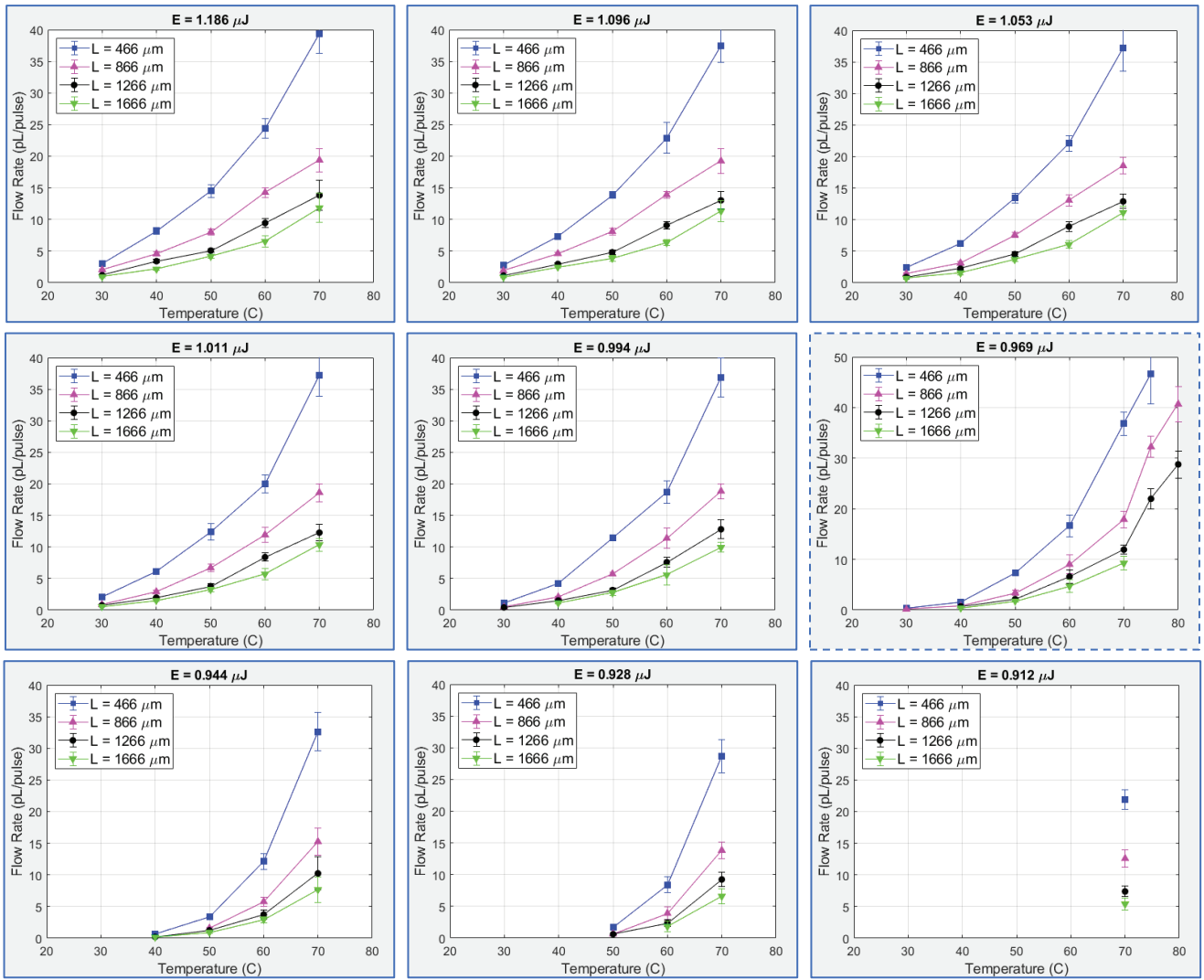}
\caption{Inertial flow rates in pL/pulse vs. operating temperature $T$. Each panel corresponds to a particular resistor energy and multiple curves within a panel correspond to different channel lengths. The symbols are mean experimental values, and error bars are three times the standard deviation. The panel with a dashed outline ($E = 0.969$ $\mu$J) also includes data for $T = 75\,^{\circ}\!{\rm C}$ and $80\,^{\circ}\!{\rm C}$. The lines are guides to the eye only.}   
\label{tdep:fig:five}
\end{figure*}

To obtain a sample flow profile, the set of tracks was split into 12-15 bins according to the tracks' $x$ position in the channel, Fig.~\ref{tdep:fig:three}(c). Within each bin, a maximal sample velocity $v_{p}(x_k)$ was taken as an estimator of the true velocity profile at an $x_k$ corresponding to the bin's midpoint. Then, the sample flow profile $\{ v_{p}(x_k) \}$ was fitted to the theoretical shape of Eqs.~(\ref{flnet:eq:atwentyfourthree}) and (\ref{flnet:eq:atwentyfourfour}) using the $x$ coordinates of the two zeros and the height as three adjustable parameters, Fig.~\ref{tdep:fig:three}(d). A standard Matlab function could accomplish a three-parameter optimization in about 1 s. Once the $v_{\rm max}$ was determined from the fit, the overall flow rate was computed from Eq.~(\ref{tdep:eq:one}).   

Finally, standard deviations of flow rates were computed by resampling the tracks at a 50\% sub-sample ratio, Fig.~\ref{tdep:fig:three}(b). For example, from a sample of 300 tracks, 150 tracks were selected at random, and a sub-sample flow rate was computed using the procedure described above. The process was then repeated 200 times resulting in a distribution of flow rates. The mean value of the distribution was taken as the final flow rate estimate for the given experimental conditions. Half-width of the distribution was taken as flow rate's standard deviation, Fig.~\ref{tdep:fig:three}(e). The error bars shown in subsequent figures are three-times the standard deviation obtained with the resampling method.

\begin{figure*}[t]
\includegraphics[width=0.98\textwidth]{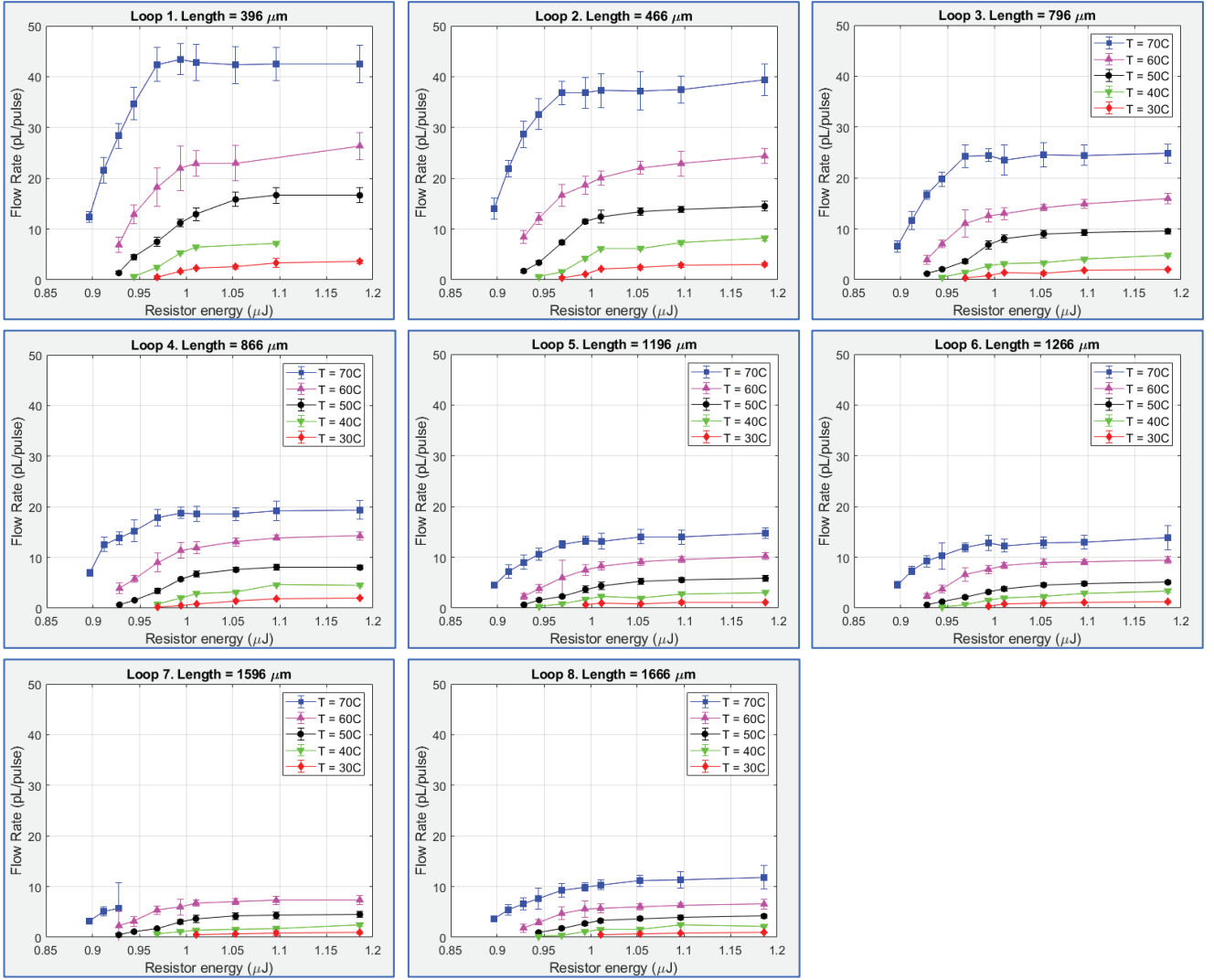}
\caption{Inertial flow rates in pL/pulse vs. resistor energy $E$. Each panel corresponds to a particular channel length and multiple curves within a panel correspond to different temperatures. The symbols are mean experimental values and error bars are three times the standard deviation. The lines are guides to the eye only.}   
\label{tdep:fig:six}
\end{figure*}

\section{\label{tdep:sec:three}
Results 
}

In this study, we measured inertial flow rates for different channel lengths, operating temperatures, and resistor energies. There are multiple ways of presenting flow rates vs three parameters. First, we show flow rate variation with channel length $L$ in Fig.~\ref{tdep:fig:four}. Each panel corresponds to a fixed resistor energy and each line within a panel corresponds to a fixed temperature. The dashed lines are best fits to the 1D model of inertial pumping.~\cite{Kornilovitch2013} In our experimental conditions, net pumping effect is dominated by a post-collapse flow, which is inversely proportional to total fluidic resistance, as explained in Sec.~\ref{tdep:sec:four}. Since fluidic resistance grows linearly with $L$, it explains the observed $1/L$ variation of flow rates. Sharp flow rate increases with temperature are also apparent in the data.

\begin{figure*}[t]
\includegraphics[width=0.98\textwidth]{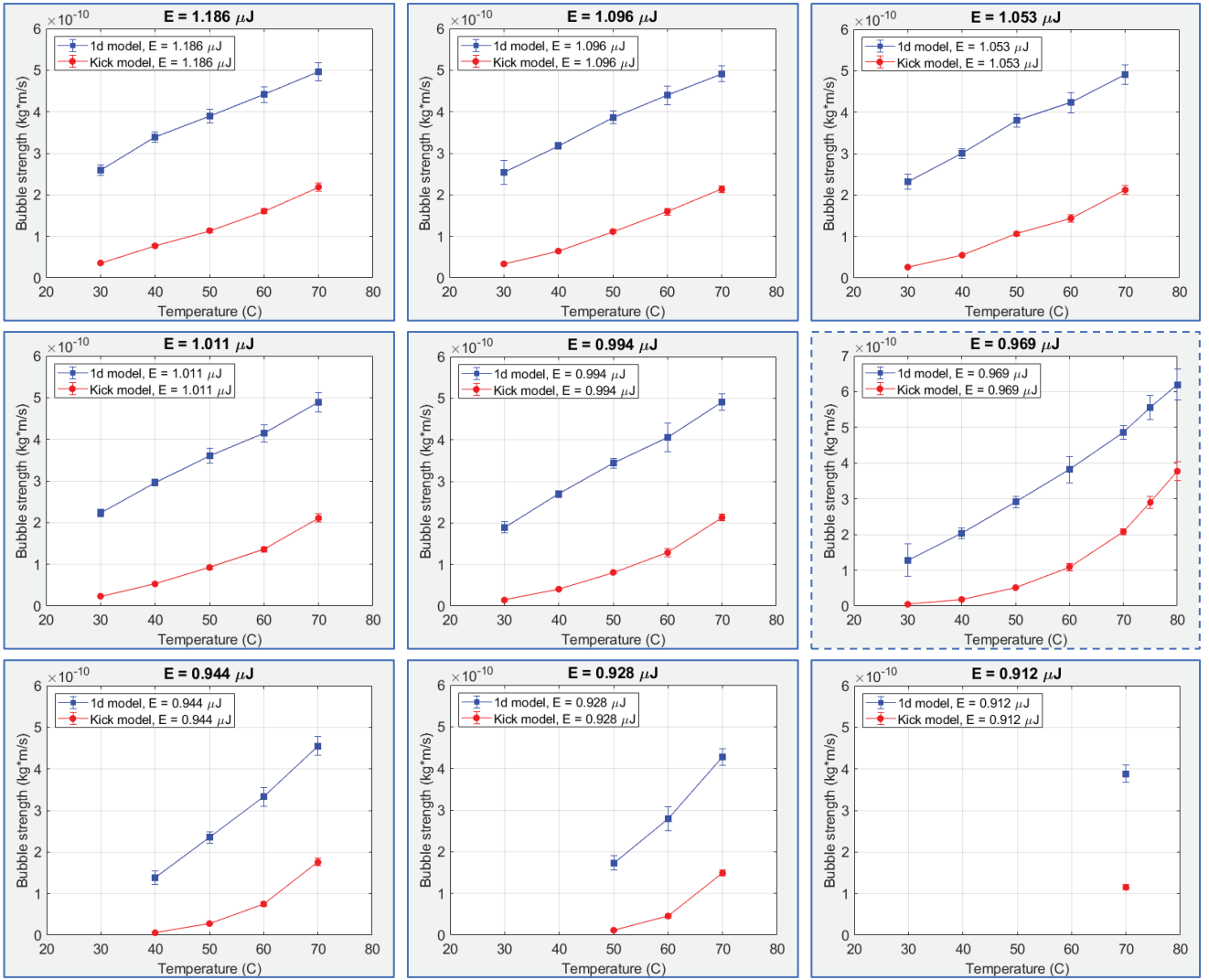}
\caption{Top lines: intrinsic bubble strength $q_0$ extracted from experimental data by fitting the $L$ dependent flow rates of Fig.~\ref{tdep:fig:four} to the 1D model of inertial pumping.~\cite{Kornilovitch2013} Bottom lines: kick strength $q$ extracted from experimental data by fitting the $L$ dependent flow rates of Fig.~\ref{tdep:fig:four} to the kick formula, Eq.~(\ref{tdep:eq:five}). }   
\label{tdep:fig:seven}
\end{figure*}

Next, we show the $T$-dependence of flow rates in Fig.~\ref{tdep:fig:five}. Steep growth is observed for all resistor energies and all channel lengths. For example, at $E = 1.186$ $\mu$J, the ratio of flow rates at $T = 70\,^{\circ}\!{\rm C}$ and $T = 30\,^{\circ}\!{\rm C}$ is about a factor of 12. For one particular energy, $E = 0.969$ $\mu$J, we investigated even higher temperatures and found a further increase up to at least $80\,^{\circ}\!{\rm C}$. Beyond $80\,^{\circ}\!{\rm C}$ inertial pumping becomes less stable due to overheating and formation of air bubbles mixed with water vapor which complicates measurement and analysis. There are three physical effects that contribute to high flow rates at elevated temperatures. First, it is the decreasing viscosity of the test fluid. The dynamic viscosity of water at $70\,^{\circ}\!{\rm C}$ is one half of that at $30\,^{\circ}\!{\rm C}$: $\eta = 0.40$ vs $0.80$ mPa s.~\cite{AntonPaar} A smaller viscosity decreases fluidic resistance and increases the flow rate. Thus, a factor of 2 out of 12 can be assigned to a changing viscosity. The remaining factor of 6 is attributed to the pump getting stronger with temperature. The latter effect can be further subdivided into two factors. First, the drive vapor bubble gets stronger. As the fluid is preheated, a progressively smaller fraction of energy released in a resistor pulse must be spent to heat the fluid to the boiling temperature which leaves more energy to produce vapor. Second, conversion of the initial bubble expansion to unidirectional push, which is pump mechanical efficiency, also improves with temperature. The analysis presented in Sec.~\ref{tdep:sec:four} enables to rigorously define pump efficiency and separate the two contributions. It will be shown that the bubble strength increases by about 2$\times$ and pump efficiency by about 3$\times$ between $30\,^{\circ}\!{\rm C}$ and $70\,^{\circ}\!{\rm C}$.      

Finally, we show flow rate variation with resistor energy in Fig.~\ref{tdep:fig:six}. The data show a clear saturation behavior. The flow rates initially increase but then become independent of energy. It happens because of the vapor lock phenomenon. At a critical energy flux, the entire fluid layer in direct contact with the resistor vaporizes and isolates the hot resistor from the rest of the fluid. Any further increase in the resistor temperature does not result in additional heat transfer to the fluid. The pumping rate becomes independent of the resistor temperature and released energy.

\begin{figure*}[t]
\includegraphics[width=0.98\textwidth]{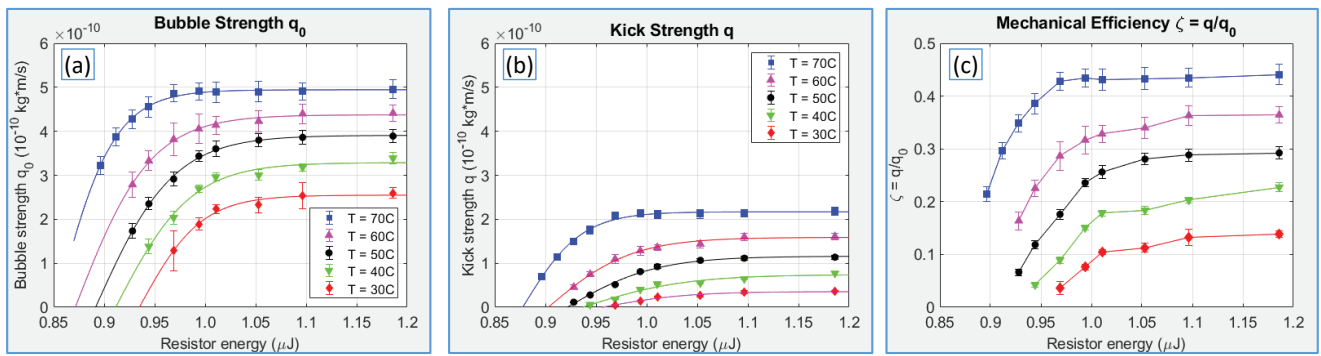}
\caption{(a) Symbols: intrinsic bubble strength $q_0$ extracted from the fits of Fig.~\ref{tdep:fig:four}, plotted vs. resistor energy and temperature. Lines: best fits to the saturation model, Eq.~(\ref{tdep:eq:seven}). (b) Symbols: kick strength $q$ obtained by fitting flow rates of Fig.~\ref{tdep:fig:four} to the kick model, Eq.~(\ref{tdep:eq:five}), plotted vs resistor energy and temperature. Lines: best fits to the saturation model, Eq.~(\ref{tdep:eq:seven}). (c) Mechanical efficiency of the inertial pump $\zeta = q/q_0$. The lines are guides to the eye only. }   
\label{tdep:fig:eight}
\end{figure*}

\section{\label{tdep:sec:four}
Analysis 
}

\subsection{\label{tdep:sec:fourone}
One-dimensional model of inertial pumping 
}

To analyze the observed dependencies, we fitted the $L$-dependent flow rates to the phenomenological 1d model of inertial pumping.~\cite{Kornilovitch2013} Not reproducing all model assumptions and equations here, it suffices to say that three-dimensional nature of the channel flow is replaced by a one-dimensional effective flow, both vapor-fluid interfaces are replaced by mathematical points, and surface tension forces are neglected compared to inertial and viscous forces. Boiling is assumed instantaneous, and resistor size is neglected relative to the channel length. The model introduces one unknown phenomenological parameter~\cite{Kornilovitch2013} --- ``bubble strength'' $q_0$ --- defined as mechanical momentum imparted on {\em both} sides of the channel at $t = 0$ and measured in (kg$\,$m/s). By order of magnitude, $q_0$ is equal to the product of peak vapor pressure in the bubble ($\sim 10^6$ Pa), time duration of the high-pressure phase ($\sim 10^{-6}$ s) and channel cross-sectional area ($\sim 10^{-9}$ m$^2$), providing an estimate $q_0 = 10^{-10} - 10^{-9}$ kg$\,$m/s. All other system's parameters including the starting expansion point (middle of the resistor), channel length, fluid viscosity (taken from tables), and fluidic resistances of moving fluidic columns as functions of interface positions, are all assumed to be independently known.       

With only one unknown parameter, in theory a single flow rate measurement is sufficient to determine $q_0$ from experimental data. In practice, this approach leads to very noisy numbers as any experimental uncertainty directly translates into a corresponding uncertainty in $q_0$. In this work, we assume that drive vapor bubbles are {\em local}. That is, the intrinsic bubble strength $q_0$ depends only on the resistor size and thermal properties of the test fluid but is independent of the channel length and other parts of the system away from the resistor. Flow rates in the eight channels should be describable by one bubble strength. In other words, the eight flow rates provide eight independent measurements of the same unknown parameter. By fitting the entire $L$-dependence to the 1d model cancels experimental noise of individual measurements and produces a more stable estimate of $q_0$. 

Such $L$ fits are shown in Fig.~\ref{tdep:fig:four} as dashed lines with best-fit $q_0$ values listed in the legends. The same values are also plotted in Fig.~\ref{tdep:fig:seven} vs temperature and in Fig.~\ref{tdep:fig:eight}(a) vs resistor energy. In can be deduced from the high-energy panels of Fig.~\ref{tdep:fig:seven} ($1.011 - 1.186$ $\mu$J) that the strength of well-developed drive bubbles increases from about $2.5\times10^{-10}$ kg$\,$m/s to about $5.0\times10^{-10}$ kg$\,$m/s between $30\,^{\circ}\!{\rm C}$ and $70\,^{\circ}\!{\rm C}$. Thus, the bubble doubles in strength between the two temperatures.

\subsection{\label{tdep:sec:fourtwo}
Simplified kick model of inertial pumping and the pump mechanical efficiency
}

The physical mechanism behind inertial pumping is complex.~\cite{Torniainen2012,Kornilovitch2013,Govyadinov2016} A key role is played by the reservoirs that dissipate fluid's momentum and provide quick refill on the shorter side of the channel. Another important parameter is the asymmetry of resistor placement within the channel. In general, total net flow consists of two contributions: (i) primary pumping effect which is a shift of the bubble collapse point relative to the starting expansion point, and (ii) secondary pumping effect caused by unequal momenta of two colliding flows at bubble collapse and resulting post-collapse flow. The full 3D CFD model and the 1d effective model both capture all those effects well.~\cite{Govyadinov2016}   

\begin{figure}[t]
\includegraphics[width=0.48\textwidth]{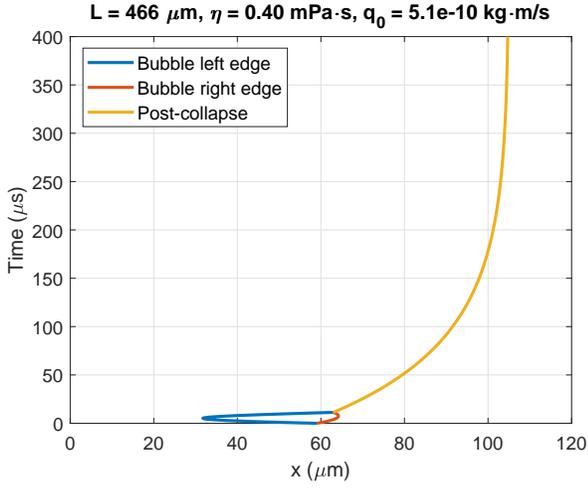}
\caption{Solutions of the one-dimensional inertial pumping equations~\cite{Kornilovitch2013} for loop 2 and $T = 70\,^{\circ}\!{\rm C}$. Notice how the primary priming effect --- shift of the bubble collapse point relative to the starting expansion point ($x = 63$ $\mu$m vs 60 $\mu$m) --- is dominated by the post-collapse flow. The overall pump dynamics can be accurately described by Poiseuille flow after an initial kick.}   
\label{tdep:fig:nine}
\end{figure}

However, in {\em effective} inertial pumps the situation is simpler. In effective pumps, the resistor is located much closer to one reservoir that to another. The shorter side of the channel empties and refills very quickly so that during that time the longer side barely moves due to its large mass and low starting velocity. As a result, the primary pumping effect (shift of the bubble collapse point relative to the resistor center) is relatively small. If the channel also has a large cross section and low fluidic resistance, post-collapse flow will persist for a long time, amplifying the secondary pumping effect. Under these conditions, which are satisfied in our system, the initial expansion-collapse phase can be neglected, in both time duration and pumping effect, compared to the post-collapse flow. The situation is illustrated in Fig.~\ref{tdep:fig:nine}. The bubble collapses at 11.3 $\mu$s and the primary (linear) pumping effect is 3.0 $\mu$m. After that, the post-collapse flow persists for another 400 $\mu$s resulting in a secondary displacement of 42 $\mu$m. One can see that with a high degree of accuracy, the entire expansion-collapse phase can be replaced by a nonzero mechanical momentum that the fluid receives when the bubble collapses. Thus, a pump pulse can be thought to result from a kick the fluid receives at time zero and then proceeds by inertia and decelerates by friction. We will be calling this simplified picture of inertial pumping the kick model. Its main advantage is extreme mathematical simplicity. Indeed, the post-collapse flow is one with a constant fluid mass (there is no vapor bubble anymore), so all the complications and nonlinearities associated with variable mass motion~\cite{Kornilovitch2013} disappear. We now derive the main equation of the kick model.

Let $q$ be the mechanical momentum imparted on the {\em entire} fluid at bubble collapse. The fluid's mass is $M = \rho A L$, so the initial velocity is $v_0 = q/M$. The friction force acting on the fluid is $\kappa \eta v L$, where $\kappa$ is the dimensionless geometric friction factor, see Appendix, Eqs.~(\ref{flnet:eq:atwentyone}) and (\ref{flnet:eq:atwentyfourfive}). $\kappa_{\rm cyl} = 8\pi$ for cylindrical channels, $\kappa_{\rm sq} = 28.45$ for 1:1 square channels, and $\kappa_{\rm sq} = 28.47$ for 30:31 channels studied in this work. Neglecting the Bernoulli correction,~\cite{Kornilovitch2013} the equation of motion reads
\begin{equation}
\rho A L \ddot{x} + \kappa \eta L \dot{x} = 0 \: ,
\label{tdep:eq:two}     
\end{equation}
where $x(t)$ is the position of a reference point and the dot denotes a derivative over time. Solving for the 
velocity and displacement one obtains
\begin{eqnarray}
v(t)       & = & v_0 \, e^{-t/\tau} \: ,
\label{tdep:eq:three} \\
x(t) - x_0 & = & v_0 \tau \left( 1 - e^{-t/\tau} \right) \: ,
\label{tdep:eq:four} 
\end{eqnarray}
where $\tau = (\rho A)/(\kappa \eta)$ is the characteristic time constant of the post-collapse flow. In our system, $\tau \sim 50-100$ $\mu$s, depending on temperature. Thus, the net flow at the end of a pulse is simply
\begin{equation}
Q = \triangle x(\infty) A = v_0 \tau A = \frac{q A}{\kappa \eta L} \: .
\label{tdep:eq:five}     
\end{equation}
This formula embodies a very transparent physics: The flow rate is kick momentum per unit cross-sectional area $q/A$ divided by the channel's fluidic resistance $\kappa \eta L/A^2$. Note that the kick model predicts a precise $1/L$ length dependence of flow rates, which is consistent with the experimental data of Fig.~\ref{tdep:fig:four}. 

{\em Kick strength} $q$ has a similar physical meaning as bubble strength $q_0$. The difference is that $q_0$ is isotropic momentum with which the original drive bubble pushes two columns of fluid in opposite directions, whereas $q$ is a {\em unidirectional} momentum acquired by the entire fluid as result of inelastic collision of two flows at bubble collapse. Since both $q$ and $q_0$ have the same physical units, their dimensionless ratio can be interpreted as the pump's {\em mechanical efficiency} $\zeta$,
\begin{equation}
\zeta = \frac{q}{q_0} \: , \hspace{1.0cm} 0 \leq \zeta \leq 1 \: . 
\label{tdep:eq:six}     
\end{equation}
$\zeta$ describes how isotropic momentum of an original drive bubble is converted to ``useful'' unidirectional kick momentum at bubble collapse.  

To determine $q$ and $\zeta$, we fitted the experimental flow rates of Fig.~\ref{tdep:fig:four} to the kick formula, Eq.~(\ref{tdep:eq:five}). Best-fit $q$ values are plotted in Fig.~\ref{tdep:fig:seven} alongside $q_0$. Focusing again on well-developed bubbles ($1.011 - 1.186$ $\mu$J), the kick strength grows from about $0.3\times10^{-10}$ to $2.0\times10^{-10}$ kg$\,$m/s between $30\,^{\circ}\!{\rm C}$ and $70\,^{\circ}\!{\rm C}$. Pump efficiency $\zeta$ increases by about 3$\times$ over the same temperature interval. The kick strength and efficiency are also plotted vs. resistor energy in Figs.~\ref{tdep:fig:eight}(b) and \ref{tdep:fig:eight}(c).

\subsection{\label{tdep:sec:fourthree}
Threshold model of energy dependence 
}

Next, we discuss the resistor energy dependence of bubble strength $q_0$ and kick strength $q$. Experimental data are shown in Fig.~\ref{tdep:fig:eight}. Both quantities display a clear saturation behavior similar to that of flow rates, cf. Fig.~\ref{tdep:fig:six}. For quantitative analysis, we fitted $q_0(E)$ and $q(E)$ to a three-parameter phenomenological threshold model
\begin{equation}
q_0(E) , \; q(E) = S \cdot \tanh{\left( \frac{ E - E_{T} }{ E_w } \right) } . 
\label{tdep:eq:seven}     
\end{equation}
Best-fit values of scale $S$, threshold energy $E_{T}$ and transition width $E_{w}$ are listed in Table~\ref{tdep:tab:two}. The resulting models are plotted in Figs.~\ref{tdep:fig:eight}(a) and \ref{tdep:fig:eight}(b) on top of experimental data.

\begin{table}[t]
\renewcommand{\tabcolsep}{0.2cm}
\renewcommand{\arraystretch}{1.5}
\begin{tabular}{|c||c|c|c|c|}
\hline\hline
    Quantity           & T, $^{\circ}$C & $S$, e-10 kg$\cdot$m/s  & $E_{T}$, $\mu$J & $E_{w}$, $\mu$J   \\ 
\hline\hline 
\multirow{5}{*}{$q_0$} &     30         &    2.548                &    0.935        &    0.0609         \\ 
                       &     40         &    3.288                &    0.911        &    0.0747         \\ 
                       &     50         &    3.908                &    0.892        &    0.0767         \\ 
                       &     60         &    4.375                &    0.871        &    0.0741         \\  
                       &     70         &    4.943                &    0.852        &    0.0567         \\ \hline
\multirow{5}{*}{$q$}   &     30         &    0.355                &    0.956        &    0.0831         \\  
                       &     40         &    0.742                &    0.937        &    0.1006         \\ 
                       &     50         &    1.160                &    0.922        &    0.0870         \\  
                       &     60         &    1.589                &    0.903        &    0.0818         \\ 
                       &     70         &    2.170                &    0.878        &    0.0575         \\ \hline 
\hline 
\end{tabular}
\caption{
Parameters of the threshold model, Eq.~(\ref{tdep:eq:seven}), for different temperatures.  
} 
\label{tdep:tab:two}
\end{table}
\begin{figure*}[t]
\includegraphics[width=0.98\textwidth]{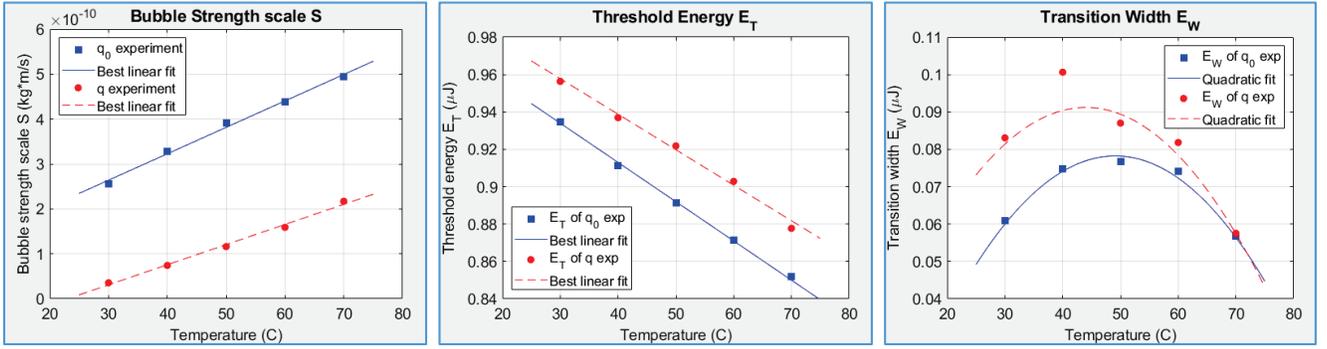}
\caption{Parameters of the threshold model, Eq.~(\ref{tdep:eq:seven}) and Table~\ref{tdep:tab:two}, plotted vs operating pump temperature. The $T = 40\,^{\circ}\!{\rm C}$ point was excluded from fitting $E_W(T)$ for the kick strength $q$. }   
\label{tdep:fig:ten}
\end{figure*}

Finally, we discuss variation of $S$, $E_{T}$, and $E_{w}$ with temperature. The values of Table~\ref{tdep:tab:two} are plotted in Fig.~\ref{tdep:fig:ten}. The scale parameter $S$ shows a clear linear growth with $T$. This is a manifestation of the bubble getting stronger with temperature, as discussed earlier. The threshold energy $E_{T}$ also shows a linear temperature dependence but with a negative slope. At higher $T$, the drive bubbles reach a saturated state at an earlier energy because less energy is spent on warming the fluid before boiling. This explains the observed trend. Finally, the width parameter $E_{w}$ shows a non-monotonic behavior with a maximum near $T = 50\,^{\circ}\!{\rm C}$. We do not have an explanation for such a behavior. 

The simplicity of the kick model allows for construction of a complete predictive flow rate model for our microfluidic system. Combining the best linear fits for $S(T)$ and $E_{T}(T)$, the best quadratic fit for $E_{w}(T)$, cf. Fig.~\ref{tdep:fig:ten}, the energy model for the kick strength $q(E)$, Eq.~(\ref{tdep:eq:seven}), and the kick formula, Eq.~(\ref{tdep:eq:five}), one obtains:
\begin{equation}
Q(T,E,L) = \frac{S(T) A}{\kappa \eta(T) L} \tanh{\left( \frac{ E - E_{T}(T) }{ E_w(T) } \right) } ,
\label{tdep:eq:eight}     
\end{equation}
where
\begin{eqnarray}
S(T)     & = & (   0.0448 \, T - 1.0351 ) \times 10^{-10} \; {\rm kg}\cdot{\rm m/s} \: ,
\label{tdep:eq:nine} \\
E_{T}(T) & = & ( - 0.0019 \, T + 1.0148 ) \; \mu{\rm J} \: ,
\label{tdep:eq:ten}  \\
E_{w}(T) & = & [ - 0.00005 \, ( T - 50 )^2 - 0.0006 \, ( T - 50 ) 
\nonumber            \\
         &   &  + 0.0894 ] \; \mu{\rm J} \: ,
\label{tdep:eq:eleven}
\end{eqnarray}
and temperature $T$ is specified in degrees $^{\circ}\!{\rm C}$. Note, however, that this model is specific to the particular resistors size, channel cross-sectional area and test fluid investigated in this work. For a different system, the model would have to be rebuilt using the methodology developed herein.

\section{\label{tdep:sec:five}
Summary and conclusions  
}

In this paper, we systematically investigated inertial pump flow rates versus channel length $L$, operating temperature $T$, and TIJ resistor thermal release energy $E$. We developed a robust methodology of measuring flow rates based on identifying the fastest tracer particle and theoretical flow profiles in rectangular channels, see Sec.~\ref{tdep:sec:twothree} and the Appendix. 

Our experimental results are summarized in Figs.~\ref{tdep:fig:four}-\ref{tdep:fig:six}. The length dependence of flow rates follows a close-to-perfect $1/L$ behavior, which is consistent with the dominant contribution of post-collapse phase to overall net flow. The most interesting findings are associated with temperature variation. We determined that flow rates delivered by the same pump can grow by as much as a factor of 12 between $T = 30\,^{\circ}\!{\rm C}$ and $T = 70\,^{\circ}\!{\rm C}$. Only partially such an increase can be accounted for by a lower viscosity while the rest must be attributed to the pump getting stronger with temperature. Through the phenomenological analysis of Sec.~\ref{tdep:sec:fourone}, we established that the drive bubble grows stronger by about 2$\times$ in the same temperature interval. The remaining contribution comes from an increasing {\em mechanical efficiency} of the pump. The latter term was introduced in Sec.~\ref{tdep:sec:fourtwo} as a consequence of a newly developed kick model of inertial pumping. It is possible that the efficiency enhancement is caused by a dropping viscosity in its own way, making the overall viscosity impact on flow rates nonlinear. This topic deserves further analysis. 

We also observed a clear saturation behavior in the resistor energy dependence of flow rates. Saturation can be understood based on vapor lock effect that limits heat flux from a TIJ resistor beyond a certain threshold. Through phenomenological modeling of Sec.~\ref{tdep:sec:fourthree}, we determined threshold and transition width parameters and established their variation with temperature. Finally, we constructed a fully predictive model of our inertial pumps in Eqs.~(\ref{tdep:eq:eight})-(\ref{tdep:eq:eleven}).      

Integrated inertial pumps are expected to become key drivers of complex microfluidic chips in the future.~\cite{Hayes2018,Govyadinov2015} Temperature variations are often an essential part of the workflows, for example, in chip-scale Polymerase Chain Reaction thermocyclers. Our results demonstrate that such system must be designed with care, as pumping rates might vary significantly depending on what part of the workflow the device is executing. Additionally, they point to the need of accurate flowmeters and temperature sensors that could ensure stable flow rates via close-loop control systems. Based on our results, we also expect significant temperature dependence in other microfluidic operations enabled by TIJ resistors, for example in active mixing.~\cite{Hayes2018b}

\begin{acknowledgments}

The authors wish to thank Carson Denison, Mandana Hamidi Haines, Brandon Hayes, Anand Jebakumar, Richard Seaver, Erik Torniainen for discussions on the subject of this paper as well as Arun Agarwal, Paul Benning and Norman Pawlowski for supporting this work.   

\end{acknowledgments}

\appendix

\section{ \label{tdep:sec:appa}
Poiseuille flow in rectangular microchannels   
}

\begin{figure*}[t]
\includegraphics[width=0.98\textwidth]{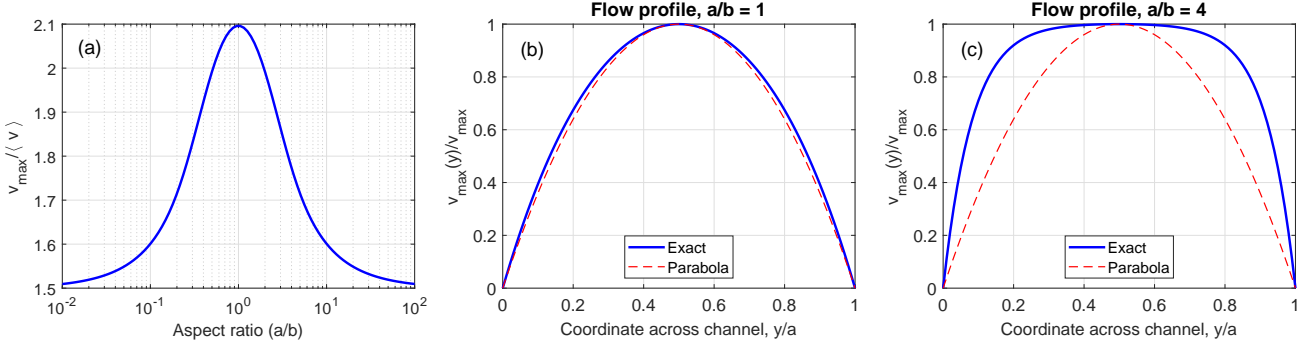}
\vspace{0.2cm}
\caption{(a) Maximal fluid velocity in the center of a rectangular channel as a function of the aspect ratio. Specific values of $v_{\rm max}/\langle v \rangle$ for (1:1) $ = 2.096$; (2:1) $ = 1.992$; (3:1) $ = 1.864$; and (10:1) $ = 1.601$. (b) Velocity profile in channels with square cross-sections, $a = b$. The red dashed line is a parabola. The thick blue line is the exact theoretical formula, Eq.~(\ref{flnet:eq:atwentyfourthree}). (c) Velocity profile in a wide-and-thin channel, $a = 4b$. The red dashed line is a parabola. The thick blue line is the exact theoretical formula, Eq.~(\ref{flnet:eq:atwentyfourthree}). Notice significant deviation from the parabolic shape and flattening of the profile in the center of the channel.}   
\label{flnet:fig:eleven}
\end{figure*}

For completeness, we derive here an analytical formula for the flow profile in rectangular channels. Pressure-driven Poiseuille flow field $v_y(x,z)$ in a channel of constant cross section satisfies a two-dimensional Poisson equation~\cite{Landau_v6,Bruus2008}   
\begin{equation}
\frac{\partial^2 v_y}{\partial x^2} + \frac{\partial^2 v_y}{\partial z^2} = 
- \frac{\triangle p}{\eta L} \: ,
\label{flnet:eq:osix}
\end{equation}
where $L$ is the channel length and $\eta$ is fluid's dynamic viscosity. The channel is oriented along the $y$ axis and the cross section lies within the $xz$ plane. $v_y = 0$ at the channel walls due to the no-slip condition. Note that the problem is mathematically equivalent to torsion of thin long elastic beams.~\cite{Timoshenko1970} Consider a rectangular channel of width $a$, $0 \leq x \leq a$, and height $b$, $0 \leq z \leq b$. Separating variables, the solution of (\ref{flnet:eq:osix}) is given by the double series:
\begin{equation}
v_y(x,z) = \frac{\triangle p}{\eta L} \frac{16}{ab} 
\sum^{\infty}_{n,m = 0} \frac{ \sin(p_{2n+1}x) \sin(q_{2m+1}z) }
{ p_{2n+1} q_{2m+1} ( p^2_{2n+1} + q^2_{2m+1} ) } \: ,
\label{flnet:eq:afourteen}
\end{equation}
\begin{equation}
p_{2n+1} = \frac{\pi ( 2n + 1 )}{a} , \hspace{0.5cm} 
q_{2m+1} = \frac{\pi ( 2m + 1 )}{b} \: .
\label{flnet:eq:aseventeen}
\end{equation}
Integral of $v_y(x,z)$ over the cross section is equal to volumetric flow rate $Q$, which in turn defines {\em average} velocity $\langle v \rangle$ as
\begin{equation}
\langle v \rangle \equiv \frac{Q}{ab} = \frac{1}{ab} \int dx \, dz \: v_{y}(x,z) \: .
\label{tdep:eq:oone}
\end{equation}
Performing integration and partial summation~\cite{Prudnikov1986} of Eq.~(\ref{flnet:eq:afourteen}), the pressure gradient can be eliminated in favor of $\langle v \rangle$:  
\begin{equation}
v_y = \langle v \rangle \frac{a b}{4 S_1}  
\sum^{\infty}_{n,m = 0} \frac{ \sin(p_{2n+1}x) \sin(q_{2m+1}z) }
{ p_{2n+1} q_{2m+1} ( p^2_{2n+1} + q^2_{2m+1} ) } \: .
\label{flnet:eq:aeighteen}
\end{equation}
Here $S_1$ is a single series that can be expressed in two equivalent forms  
\begin{eqnarray}
S_1(a,b) & = & \frac{a^2 b^4}{3 \cdot 2^8} 
\left\{ 1 - \frac{192}{\pi^5} \left( \frac{b}{a} \right) \cdot S_{3m}(a,b) \right\}
\label{flnet:eq:aseventeenone} \\
         & = & \frac{a^4 b^2}{3 \cdot 2^8} 
\left\{ 1 - \frac{192}{\pi^5} \left( \frac{a}{b} \right) \cdot S_{3n}(a,b) \right\} ,
\label{flnet:eq:aseventeentwo}
\end{eqnarray}
\begin{equation}
S_{3m}(a,b) = \sum^{\infty}_{m = 0} \frac{ 1 } { ( 2m+1 )^5 } \tanh{ \frac{\pi a (2m+1) }{2b} } \: ,
\label{flnet:eq:aseventeenthree}
\end{equation}
\begin{equation}
S_{3n}(a,b) = \sum^{\infty}_{n = 0} \frac{ 1 } { ( 2n+1 )^5 } \tanh{ \frac{\pi b (2n+1) }{2a} } \: .
\label{flnet:eq:aseventeenfour}
\end{equation}

The ``flow profile'' method of Sec.~\ref{tdep:sec:twothree} requires knowledge of the {\em maximal fluid velocity for a given} $x$. It is clear from symmetry considerations that a maximum velocity is reached at $z = b/2$ for any $x$. Setting $z = b/2$ in Eq.~(\ref{flnet:eq:aeighteen}) yields
\begin{equation}
v_{\rm m}(x) = \langle v \rangle \frac{a b}{4 S_1}  
\sum^{\infty}_{n,m = 0} \frac{ \sin(p_{2n+1}x) (-1)^m }
{ p_{2n+1} q_{2m+1} ( p^2_{2n+1} + q^2_{2m+1} ) } \: .
\label{flnet:eq:atwentythreeseven}
\end{equation}
The sum over $m$ can be calculated analytically~\cite{Prudnikov1986}
\begin{equation}
v_{\rm m}(x) = \langle v \rangle \frac{a b^2}{16 S_1}  
\sum^{\infty}_{n = 0} \frac{ \sin(p_{2n+1}x) }{ p^3_{2n+1} } 
\left( 1 - \frac{1}{\cosh{\frac{p_{2n+1} b}{2}} } \right) .
\label{flnet:eq:atwentythreeeight}
\end{equation}
Out of the remaining two sums, the first one can also be calculated analytically,~\cite{Prudnikov1986} yielding the flow profile
\begin{equation}
v_{\rm m}(x) = \langle v \rangle \frac{a^2 b^2}{2^7 S_1}  
\left\{ x ( a - x ) - \frac{8}{a} 
\sum^{\infty}_{n = 0} \frac{ \sin(p_{2n+1}x) }{ p^3_{2n+1} \cosh{\frac{p_{2n+1} b}{2}} } 
\right\} \! .
\label{flnet:eq:atwentythreenine}
\end{equation}

Next, we calculate the {\em absolute} maximum velocity, which is reached in the middle of a channel. Setting $x = a/2$ in Eq.~(\ref{flnet:eq:atwentythreenine}), one obtains  
\begin{equation}
v_{\rm max}(a,b) = \langle v \rangle \cdot \frac{3}{2}  
\frac{ 1 - \frac{32}{\pi^3} \cdot S_{4n}(a,b) }
     { 1 - \frac{192}{\pi^5} \left( \frac{a}{b} \right) \cdot S_{3n}(a,b) } \: ,
\label{flnet:eq:atwentyfourone}
\end{equation}
where
\begin{equation}
S_{4n}(a,b) \equiv \sum^{\infty}_{n = 0} 
\frac{(-1)^n}{( 2n + 1 )^3 \cosh{\frac{\pi b ( 2n + 1 ) }{2a}}} \: ,
\label{flnet:eq:atwentyfourtwo}
\end{equation}
and $S_{3n}$ is defined in Eq.~(\ref{flnet:eq:aseventeenfour}). Note that $v_{\rm max}$ can also be expressed in an equivalent form 
\begin{equation}
v_{\rm max}(a,b) = \langle v \rangle \cdot \frac{3}{2} 
\frac{ 1 - \frac{32}{\pi^3} \cdot S_{4m}(a,b) }
     { 1 - \frac{192}{\pi^5} \left( \frac{b}{a} \right) \cdot S_{3m}(a,b) } \: ,
\label{flnet:eq:atwentythreeone}
\end{equation}
\begin{equation}
S_{4m}(a,b) = \sum^{\infty}_{m = 0} 
\frac{(-1)^m}{( 2m + 1 )^3 \cosh{\frac{\pi a ( 2m + 1 ) }{2b}}} \: ,
\label{flnet:eq:atwentythreetwo}
\end{equation}
which reflects the fact that the maximum-to-average velocity ratio is invariant under transformation $a \leftrightarrow b$ that leaves the channel's aspect ratio unchanged. It is possible to show that in both limits $a \gg b$ and $a \ll b$, $v_{\rm max} \rightarrow \frac{3}{2} \langle v \rangle$, the value for an infinite slot. The full Eqs.~(\ref{flnet:eq:atwentyfourone}) and (\ref{flnet:eq:atwentythreeone}) vs aspect ratio $(a/b)$ is shown in Fig.~\ref{flnet:fig:eleven}(a).  

Returning now to the task of calculating the velocity profile, divide Eq.~(\ref{flnet:eq:atwentythreenine}) by Eq.~(\ref{flnet:eq:atwentyfourone}) to get, after transformations
\begin{equation}
v_{\rm max}(x) = v_{\rm max} \cdot  
\frac{ 4 \left( \frac{x}{a} \right) \left( 1 - \frac{x}{a} \right) - \frac{32}{\pi^3} \cdot S_{6}(x) }
     { 1 - \frac{32}{\pi^3} \cdot S_{4n} } \: ,
\label{flnet:eq:atwentyfourthree}
\end{equation}
\begin{equation}
S_{6}(x ; a,b) \equiv \sum^{\infty}_{n = 0} 
\frac{\sin{\frac{\pi x}{a}( 2n + 1 )}}{( 2n + 1 )^3 \cosh{\frac{\pi b ( 2n + 1 ) }{2a}}} \: .
\label{flnet:eq:atwentyfourfour}
\end{equation}
The flow profile, Eq.~(\ref{flnet:eq:atwentyfourthree}), is the sum of a parabola plus a linear combination of sine harmonics. The relative contribution of the sines depends on the channel's aspect ratio. In tall and skinny channels, $b \gg a$, the sines' weights are exponentially small, and the flow profile is a parabola. As $a/b$ increases, the sine weights grow to be non-negligible and in square channels, $a = b$, deviation from a pure parabola is already noticeable although it remains small, see Fig.~\ref{flnet:fig:eleven}(b). However, in thin and wide channels, $a \gg b$, many harmonics contribute rendering the flow profile substantially non-parabolic, as illustrated in Fig.~\ref{flnet:fig:eleven}(c) for the case of $a/b = 4$.   

Calculating the viscous stress tensor 
\begin{equation}
\sigma_{ik} = \eta \left( \frac{\partial v_i}{\partial x_k} + \frac{\partial v_k}{\partial x_i} \right) ,
\label{flnet:eq:aonethree}
\end{equation}
and integrating over the perimeter of the channel cross section, the total friction force can be expressed as $F = \kappa \cdot \eta \langle v \rangle L$, where $\eta$ is the dynamic viscosity and $L$ is the channel length. The dimensionless coefficient $\kappa$ is 
\begin{equation}
\kappa_{\rm rectangle} = \frac{a^3 b^3}{64 \: S_1(a,b)} \: . 
\label{flnet:eq:anineteen}
\end{equation}
Using Eqs.~(\ref{flnet:eq:aseventeenthree}) and (\ref{flnet:eq:aseventeenfour}) yields the final result
\begin{eqnarray}
\kappa_{\rm rectangle} & = & \frac{12 a}{b} 
\frac{1}{ 1 - \frac{192}{\pi^5} \left( \frac{b}{a} \right) \cdot S_{3m}(a,b)} \: ,
\label{flnet:eq:atwentyone} \\
                       & = & \frac{12 b}{a} 
\frac{1}{ 1 - \frac{192}{\pi^5} \left( \frac{a}{b} \right) \cdot S_{3n}(a,b)} \: .
\label{flnet:eq:atwentyfourfive}
\end{eqnarray}
For the channels studied in this work, $a/b = 30/31$ and $\kappa = 28.47$.


\begin{thebibliography}{100}

\bibitem{Yuan1999}
H. Yuan and A. Prosperetti,
``The pumping effect of growing and collapsing bubbles in a tube,''
J. Micromech. Microeng. {\bf 9}, 402--413 (1999).

\bibitem{Yin2005a}
Z. Yin and A. Prosperetti,
``A microfluidic `blinking bubble' pump,''
J. Micromech. Microeng. {\bf 15}, 643--651 (2005).

\bibitem{Yin2005b}
Z. Yin and A. Prosperetti,
`` `Blinking bubble' micropump with microfabricated heaters,''
J. Micromech. Microeng. {\bf 15}, 1683--1691 (2005).

\bibitem{Torniainen2012}
E.D. Torniainen, A.N. Govyadinov, D.P. Markel, and P.E. Kornilovitch,
``Bubble\textendash driven inertial micropump,''
Phys. Fluids {\bf 24}, 122003 (2012).

\bibitem{Govyadinov2016}
A.N. Govyadinov, P.E. Kornilovitch, D.P. Markel, and E.D. Torniainen, 
``Single\textendash pulse dynamics and flow rates of inertial micropumps,''
Microfluid. Nanofluid. {\bf 20}, 73 (2016).

\bibitem{Stasiak2012}
J. Stasiak, S. Richards, and P. Benning, 
``Hewlett-Packard's MEMS technology --- Thermal inkjet printing and beyond,''
in {\em Microelectronics to Nanoelectronics: Materials, Devices \& Manufacturability}, edited by A. B. Kaul (CRC/Taylor \&
Francis, Boca Raton, FL, 2012).

\bibitem{Kornilovitch2013}
P.E. Kornilovitch, A.N. Govyadinov, D.P. Markel, and E.D. Torniainen, 
``One\textendash dimensional model of inertial pumping,''
Phys. Rev. E {\bf 87}, 023012 (2013).

\bibitem{Hayes2021}
B. Hayes, G.L. Whiting, and R. MacCurdy,
``Modeling of contactless bubble\textendash bubble interactions in microchannels with integrated inertial pumps,''
Phys. Fluids {\bf 33}, 042002 (2021). 

\bibitem{Hayes2018}
B.S. Hayes, A.N. Govyadinov, and P.E. Kornilovitch,  
``Microfluidic switchboards with integrated inertial pumps,''
Microfluid. Nanofluid. {\bf 22}, 15 (2018).

\bibitem{Wang2004}
G.R. Wang, J.G. Santiago, M.G. Mungal, B. Young, and S. Papandemetriou, 
``A laser induced cavitation pump,''
J. Micromech. Microeng. {\bf 14}, 1037--1046 (2004).

\bibitem{Dijkink2008}
R.J. Dijkink and C.D. Ohl, 
``Laser-induced cavitation based micropump,'' Lab Chip {\bf 8}, 1676--1681 (2008).

\bibitem{Schindelin2012}
J. Schindelin, I. Arganda-Carreras, E. Frise, V. Kaynig, M. Longair, T. Pietzsch, S. Preibisch, C. Rueden, S Saalfeld, B.Schmid, J.-Y. Tinevez, D.J. White, V. Hartenstein, K. Eliceiri, P. Tomancak, and A. Cardona,  
``Fiji: An open-source platform for biological-image analysis,''
Nature Methods {\bf 9}, 676--682 (2012).

\bibitem{Tinevez2017}
J.-Y. Tinevez, N. Perry, J. Schindelin, G.M. Hoopes, G.D. Reynolds, E. Laplantine, S.Y. Bednarek, S.L. Shorte, and K.W. Eliceiri,  
``TrackMate: An open and extensible platform for single-particle tracking,''
Methods {\bf 115}, 80--90 (2017).

\bibitem{Govyadinov2015}
A.N. Govyadinov, E.D. Torniainen, P.E. Kornilovitch, and D.P. Markel, 
``Path to low cost microfluidics,''
in {\em Proceedings of NIP31, Portland, OR} (IS\&T 2015), pp. 490--497.

\bibitem{AntonPaar}
See https://wiki.anton-paar.com/us-en/water/ for tables of water viscosity and density vs temperature.

\bibitem{Hayes2018b}
B. Hayes, A. Hayes, M. Rolleston, A. Ferreira, and J. Kirsher, 
``Pulsatory mixing of laminar flow using bubble-driven micro-pumps,'' 
in {\em Proceedings of the ASME 2018 International Mechanical Engineering Congress and Exposition} (ASME 2018), Vol. 7. 

\bibitem{Landau_v6}
L.D. Landau and E.M. Lifshits, 
{\em Course of Theoretical Physics, Volume 6: Fluid Mechanics}, 2nd ed. (Elsevier Ltd., 1987).

\bibitem{Bruus2008}
H. Bruus,
{\em Theoretical Microfluidics} (Oxford University Press, 2008), Chap. 3.

\bibitem{Timoshenko1970}
S.P. Timoshenko and J.N. Goodier, {\em Theory of Elasticity}, 3rd ed. (McGaw-Hill, Inc., 1970).

\bibitem{Prudnikov1986}
A.P. Prudnikov, Yu.A. Brychkov, and O.I. Marichev,
{\em Integrals and Series, Volume 1: Elementary Functions} (Taylor \& Francis), Secs. 5.1 and 5.4. 





\end{thebibliography}
\end{document}